%% file: Main.tex
\documentclass[a4paper,fleqn,usenatbib]{mnras}

\usepackage{newtxtext,newtxmath}
\usepackage[T1]{fontenc}
\usepackage{ae,aecompl}

\usepackage{graphicx}	
\usepackage{amsmath}	
\usepackage{amssymb}	
\usepackage{adjustbox}
\usepackage{xspace}
\usepackage{natbib}
\setcitestyle{notesep={}}
\usepackage{etoolbox}
\makeatletter
\patchcmd\@combinedblfloats{\box\@outputbox}{\unvbox\@outputbox}{}{%
  \errmessage{\noexpand\@combinedblfloats could not be patched}%
}%
\makeatother
\usepackage{cleveref}
\crefname{figure}{Figure}{Figures}
\crefname{section}{Section}{Sections}
\crefname{table}{Table}{Tables}

\newcommand{\Msol}{M$_{\odot}$\xspace}

\newcommand{\squotes}[1]{\lq {#1}\rq\xspace}
\newcommand{\dquotes}[1]{\lq\lq {#1}\rq\rq\xspace}
\newcommand{\eagle}{{\sc eagle}\xspace}

\newcommand{\subfind}{{\sc subfind}\xspace}

\newcommand{\M}[1]{$M_{\mathrm{#1}}$\xspace}

\newcommand{\ndyn}{$n_{\mathrm{dyn}}$\xspace}
\newcommand{\ndynx}[1]{$n_{\mathrm{dyn[#1]}}$\xspace}
\newcommand{\tmerger}{$t_{\mathrm{merger}}$\xspace}
\newcommand{\erg}{erg~s$^{-1}$\xspace}

\title[The rapid growth phase of supermassive black holes]{The rapid growth phase of supermassive black holes}

\author[S. McAlpine et al.]{Stuart McAlpine$^{1}$\thanks{E-mail:
s.r.mcalpine@durham.ac.uk}, Richard G. Bower$^{1}$, David J. Rosario$^{2}$, Robert A. Crain$^{3}$, \newauthor Joop Schaye$^{4}$, and Tom Theuns$^{1}$.
\\
$^{1}$Institute for Computational Cosmology, Department of Physics, Durham University, South Road, Durham, DH1 3LE, UK\\
$^{2}$Centre for Extragalactic
Astronomy, Department of Physics, Durham University, South Road, Durham DH1
3LE, UK\\$^{3}$Astrophysics Research Institute, Liverpool John Moores
University, 146 Brownlow Hill, Liverpool L3 5RF, UK\\
$^{4}$Leiden Observatory,
Leiden University, P.O. Box 9513, 2300 RA Leiden, the Netherlands}

\date{Accepted XXX. Received YYY; in original form ZZZ}

\pubyear{2018}

\begin{document}
\label{firstpage}
\pagerange{\pageref{firstpage}--\pageref{lastpage}}
\maketitle

\input{Results.tex}
\bibliographystyle{mnras}
\bibliography{mybib}

\bsp	
\label{lastpage}

\end{document}

%% file: Results.tex
\begin{abstract}

We investigate the rapid growth phase of supermassive black holes (BHs) within the hydrodynamical cosmological \eagle simulation. This non-linear phase of BH growth occurs within $\sim$$L_{*}$ galaxies, embedded between two regulatory states of the galaxy host: in sub $L_{*}$ galaxies efficient stellar feedback regulates the gas inflow onto the galaxy and significantly reduces the growth of the central BH, while in galaxies more massive than $L_{*}$ efficient AGN feedback regulates the gas inflow onto the galaxy and curbs further non-linear BH growth. We find evolving critical galaxy and halo mass scales at which rapid BH growth begins. Galaxies in the low-redshift Universe transition into the rapid BH growth phase in haloes that are approximately an order of magnitude more massive than their high-redshift counterparts (\M{200} $\approx 10^{12.4}$~\Msol at $z \approx 0$ decreasing to \M{200} $\approx 10^{11.2}$~\Msol at $z \approx 6$). Instead, BHs enter the rapid growth phase at a fixed critical halo virial temperature ($T_{\mathrm{vir}} \approx 10^{5.6}$~K). We additionally show that major galaxy--galaxy interactions ($\mu \geq \frac{1}{4}$, where $\mu$ is the stellar mass ratio) play a substantial role in triggering the rapid growth phase of BHs in the low-redshift Universe, whilst potentially having a lower influence at high redshift. Approximately 40\% of BHs that initiate the rapid BH growth phase at $z \approx 0$ do so within $\pm 0.5$ dynamical times of a major galaxy--galaxy merger, a fourfold increase above what is expected from the background merger rate. We find that minor mergers ($\frac{1}{10} \leq \mu < \frac{1}{4}$) have a substantially lower influence in triggering the rapid growth phase at all epochs.    
\end{abstract}

\begin{keywords}
galaxies: active -- galaxies: evolution -- galaxies: formation -- galaxies: high-redshift -- galaxies: interactions
\end{keywords}

\section{Introduction}

Feedback from star formation, including stellar winds, radiation pressure and supernovae, plays a key role in galaxy evolution. Collectively described as \squotes{stellar feedback}, the energy injection into the surrounding interstellar medium can eject material from the galaxy via an outflow \citep[see][ for a review]{Veilleux2005}. In the absence of this process, many observed phenomena within the galaxy population simply cannot be reproduced by current models: such as the relatively low percentage of baryons that eventually convert into stars \citep[$\approx 10$\%, e.g.,][]{Fukugita1998}, the flattening of the faint-end slope of the luminosity function \citep[e.g.,][]{White1978,Dekel1986,Benson2003}, the formation of exponential disks \citep[e.g.,][]{Binney2001,Scannapieco2008}, the formation of dark matter cores \citep[e.g.,][]{Navarro1996}, the cosmic star formation history \cite[e.g.,][]{White1991} and the chemical enrichment of the intergalactic medium \citep[e.g.,][]{Aguirre2001}.

At masses below $\sim$$L_*$ (\M{200} $\sim 10^{12}$~\Msol), galaxies maintain a quasi-equilibrium, with the star formation rate and the associated supernovae-driven outflow balancing the rate of the cosmic inflow \citep[e.g.,][]{White1991,Finlator2008,Bouche2010,Schaye2010}. However, as galaxies evolve past $\sim$$L_{*}$, stellar feedback becomes unable to effectively remove material from the galaxy, and the equilibrium breaks \citep[e.g.,][]{Benson2003,Hopkins2014,Keller2016}. A further source of energy is therefore required to balance against the cosmic inflow and restore the quasi-equilibrium, which is commonly attributed to the feedback from the central supermassive back hole \citep[BH, e.g.,][]{Croton2006,Bower2006,BoothandSchaye2010}.

Beyond affecting the continued production of stars within the galaxy, it is plausible that stellar feedback can also significantly hinder the growth of the central supermassive BH in sub $\sim$$L_{*}$ galaxies, where stellar feedback remains able to drive an effective outflow, and starve the inner regions of fuel for BH accretion. This result is indeed found by many current hydrodynamical simulations \citep[e.g.,][]{Dubois2015,AnglesAlcazar2017,Bower2017,Habouzit2017}. The critical point at which the stellar feedback driven outflows begin to stall will naturally be linked to the first meaningful period of BH growth. However, the critical mass scale at which this transition occurs, the triggering mechanism, and the growth of the BH during this time, remain uncertain. 

In this study we utilize the \eagle cosmological hydrodynamical simulation \citep{Schaye2015,Crain2015} to investigate the evolution of 1,888 massive BHs (\M{BH} $\geq 10^{7}$~\Msol) and the host galaxies during the rapid growth phase. This large sample of BHs allows us for the first time to link the stalling of stellar feedback driven outflows to the initiation of rapid BH growth in statistical detail, and measure the importance of external events, such as galaxy--galaxy mergers, to this period of BH evolution.
 
The paper is organized as follows. In \cref{sect:method} we briefly describe the \eagle simulations, our BH sample selection, how we define the time of the rapid growth phase and how we define the \squotes{most proximate} merger. \cref{sect:results} contains our main results, \cref{sect:discussion} outlines our discussion and in \cref{sect:conclusions} we present our conclusions.

\section{The \eagle simulation}
\label{sect:method}

The \dquotes{Evolution and Assembly of GaLaxies and their Environment}
\citep[\eagle,][]{Schaye2015,Crain2015} \footnote{\url{www.eaglesim.org}}\textsuperscript{,}\footnote{Galaxy and halo catalogues of the simulation suite, as well as the particle data, are publicly available at \url{http://www.eaglesim.org/database.php} \citep{McAlpine2015,EAGLE2017}.} is a suite of hydrodynamical cosmological simulations that cover a wide range of periodic volumes, numerical resolutions and physical models. To incorporate the processes that operate below the simulation resolution a series of \squotes{subgrid} prescriptions are implemented, namely: radiative cooling and photo-ionisation heating \citep{Wiersma2009a}; star formation \citep{Schaye2008}, stellar evolution \citep{Wiersma2009b} and stellar feedback \citep{DallaVecchia_Schaye2012}; BH growth via accretion and mergers and BH feedback \citep{Springel2005a,RosasGuevara2015}. These are calibrated to reproduce the observed galaxy stellar mass function, galaxy sizes and normalization of the BH mass--bulge mass relation at $z \approx 0.1$.  A full description of the simulation and the calibration strategy can be found in \citet{Schaye2015} and \citet{Crain2015} respectively. 

For this study we are interested in the evolution of massive BHs (\M{BH} $\geq 10^{7}$\Msol), and therefore restrict our study to the largest simulation, Ref-L0100N1504, which contains the greatest number of these objects.  This simulation is a cubic periodic volume 100 comoving megaparsecs (cMpc) on each side, containing $1504^{3}$ dark matter particles of mass $9.7 \times 10^{6}$~\Msol and an equal number of baryonic particles with an initial mass of $1.8 \times 10^{6}$~\Msol.  The subgrid parameters are those of the \eagle reference model, described fully by \citet{Schaye2015}. The cosmological parameters are those inferred by \citet{Planck2013}:
$\Omega_{\mathrm{m}}=0.307$, $\Omega_{\Lambda}=0.693$,
$\Omega_{\mathrm{b}}=0.04825$, $h = 0.6777$ and $\sigma_{8}=0.8288$.

Halo mass, \M{200}, is defined as the total mass enclosed within $r_{\mathrm{200}}$, the radius at which the mean enclosed density is 200 times the critical density of the Universe. Galaxy mass, \M{*}, is defined as the total stellar content bound to a subhalo within a spherical aperture with radius 30~proper kiloparsecs (pkpc), as per \citet{Schaye2015}.

Galaxy histories are tracked using a merger tree. As the hierarchical build-up of galaxies can be complex, the history of each galaxy is considered from the reference frame of the \squotes{main progenitor}, the branch of the galaxy's full merger tree that contains the greatest total mass \citep[see][ for details]{Qu2017}. The completion time of a galaxy--galaxy merger is defined as the cosmic time of the first simulation output where two galaxies that were previously identified as separate individually bound objects are now identified as a single bound object by the \subfind algorithm \citep{Springel2001,Dolag2009}. There are 200 simulation outputs between redshifts $z=20$ and $z=0$ at intervals of 40 to 80 Myr. Mergers are classified by the stellar mass ratio,  $\mu = M_{*,1} / M_{*,2}$, where $M_{*,2}$ is the stellar mass of the most massive member of the binary. They are considered major if $\mu \geq \frac{1}{4}$, minor if $\frac{1}{10} \leq \mu < \frac{1}{4}$ and either major or minor if $\mu \geq \frac{1}{10}$. To account for the effect of stellar stripping during the later stages of the interaction, the stellar masses are computed when the in-falling galaxy had its maximum mass \citep[e.g.,][]{Rodriguez-Gomez2015,Qu2017}. To account for the resolution of the simulation, mergers are only considered \squotes{resolved}  when $M_{*,2} \geq 10^{8}$ \Msol ($\approx 100$ stellar particles).

\subsection{The phases of black hole growth}
\label{sect:how_bhs_grow}

BHs in the \eagle simulation transition through three distinct phases of growth, governed by the mass (or more strictly the virial temperature) of the host dark matter halo. As we will repeatedly use the terminology adopted by previous studies, we briefly revisit their meaning here.  For a more comprehensive description of these phases and how they affect the observable properties of galaxies and their central BHs see \citet{McAlpine2017}, for a physical interpretation of these phases see \citet{Bower2017} \citep[see also][ for related, but different, interpretations]{Dubois2015,AnglesAlcazar2017}. 

\begin{enumerate}[I]

\item \textit{The stellar feedback regulated phase}: the buoyant outflows created by stellar feedback efficiently regulate the gas content of galaxies residing in low-mass haloes (\M{200} $\ll 10^{12}$ \Msol). As a consequence, the central density of gas in these systems remains low, resulting in only limited growth of the central BH. In this phase BHs tend to remain close to the seed mass\footnote{$M_{\mathrm{BH[seed]}} = 1.48 \times 10^{5}$ \Msol for the reference model.}.

\item \textit{The non-linear/rapid black hole growth phase}: as haloes evolve towards \M{200} $\sim 10^{12}$~\Msol the virial temperature of the halo surpasses that of the stellar outflow, causing them to stall (as they can no longer buoyantly rise). This gives the first opportunity for a high gas density to build up in the galaxy center. Now the central BH is able to grow nearly unhindered, doing so initially at a highly non-linear rate, arising since Bondi-like accretion is proportional to the mass of the BH squared \citep{Bondi1944}. We will interchangeably refer to this phase of evolution as either the  \squotes{non-linear} or \squotes{rapid growth} phase.

\item \textit{The AGN feedback regulated phase}: after the rapid growth phase, the central BH has become massive ($\gtrsim 10^{7}$ \Msol). It can now effectively regulate the gas inflow onto the halo via efficient AGN feedback. Therefore in massive haloes (\M{200} $\gtrsim 10^{12}$ \Msol) regulatory equilibrium is once again restored, and the specific growth of the BH retires to a lower rate.

\end{enumerate}

\subsection{Black hole sample selection}
\label{sect:sample_selection}

Our sample comprises all BHs more massive than $10^{7}$ \Msol at $z=0$. We only consider BHs more massive than this as they have  likely completed the non-linear phase and will have entered the AGN feedback regulated phase. This ensures that the three phases of growth outlined in \cref{sect:how_bhs_grow} can be robustly identified. A lower mass cut would contaminate the sample with a large number of BHs still undergoing the non-linear phase. We estimate this mass cut via an inspection of the BH mass--halo mass relation \citep[see Figure 2 of][]{McAlpine2017}, selecting the pivot point that marks the transition from a supra-linear to $\approx$ linear relation between the two properties. This yields a total sample of 1,888 BHs.  

\subsubsection{Identifying the non-linear phase of black hole growth}
\label{sect:identifying_nlg}

To segregate the BHs within our sample into the three evolutionary phases outlined in \cref{sect:how_bhs_grow}, we require a robust identification of the beginning and end of the non-linear phase. BHs enter the non-linear growth phase at $\approx$ the seed mass, as growth is curtailed in the preceding stellar feedback regulated phase \citep{McAlpine2017}. The specific black hole accretion rate (sBHAR\footnote{As instantaneous BH activity is highly variable \citep[see Figure 1 in][]{McAlpine2017}, the value of $\dot M_{\mathrm{BH}}$ used in all our sBHAR calculations is the 50~Myr time-averaged rate.}, the accretion rate of the BH normalized by the BH mass, i.e., $\dot M_{\mathrm{BH}} / M_{\mathrm{BH}}$) during the non-linear phase is naturally large, due to the high $\dot M_{\mathrm{BH}}$ and the relatively low $M_{\mathrm{BH}}$ over this period. Therefore, to first order, the peak of the sBHAR history provides a good estimate for when the non-linear growth phase is occurring. We then estimate the extent of the non-linear phase by tracing the log$_{10}$\M{BH} history in each direction, starting from the sBHAR peak. When the gradient, d(log$_{10}$\M{BH})/d$t$, shallows below a critical value, we take these thresholds to be the start and end points of non-linear growth, $t_\mathrm{NLG[start]}$ and $t_\mathrm{NLG[end]}$ respectively. We find the value d(log$_{10}$ \M{BH})/d$t$ = 0.25 dex~Gyr$^{-1}$ provides a robust separation of the three phases for our BH sample; however the results are insensitive to the choice of this value.  

\begin{figure}
\includegraphics[width=\columnwidth]{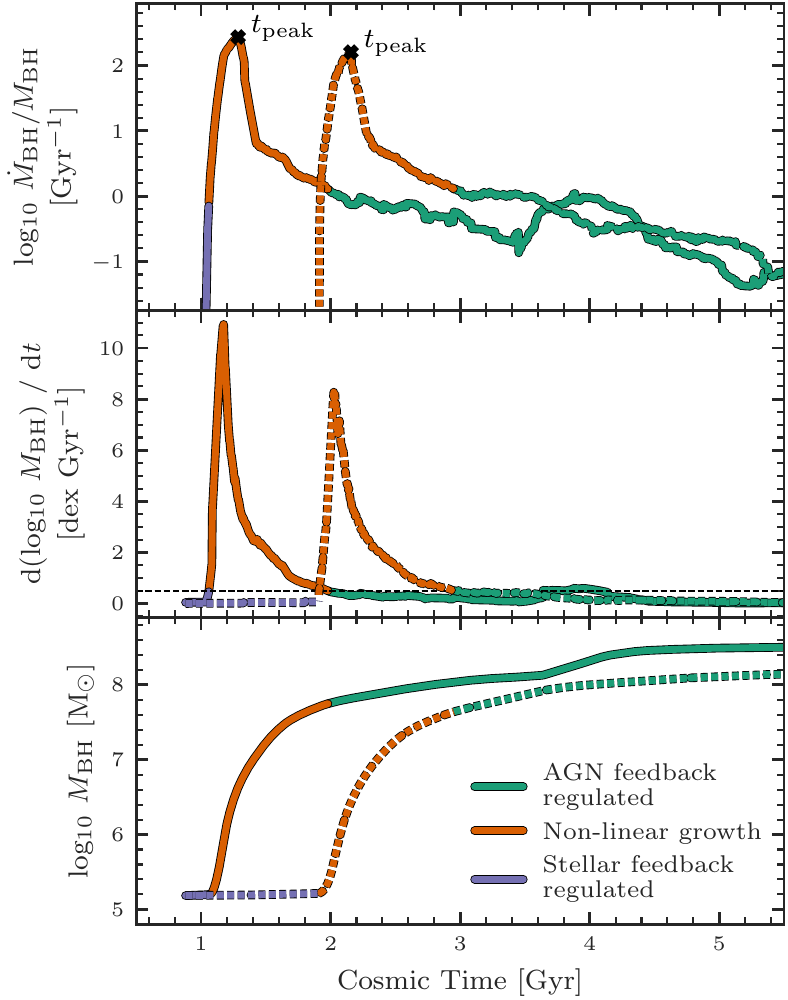}

\caption{Two illustrative examples of how the start and end points ($t_\mathrm{NLG[start]}$ and $t_\mathrm{NLG[end]}$ respectively) of the non-linear phase of BH growth are computed. Each panel is plotted as a function of cosmic time. \textit{Top panel}: the 50~Myr time-averaged sBHAR ($\dot M_{\mathrm{BH}} / M_{\mathrm{BH}}$), annotated with the maximum value, $t_\mathrm{peak}$. \textit{Middle panel}: the gradient of log$_{10}M_{\mathrm{BH}}$, d(log$_{10}$\M{BH})/d$t$. Where the gradient crosses the threshold value of d(log$_{10}$\M{BH})/d$t$ = 0.25 dex~Gyr$^{-1}$ (shown as a horizontal dashed line) in each direction, starting from $t_\mathrm{peak}$, defines the times $t_\mathrm{NLG[start]}$ and $t_\mathrm{NLG[end]}$. \textit{Bottom panel}: The BH mass. Each line is colour coded via the identified phase of BH evolution, as indicated by the legend.}

\label{fig:nlg_time_definitions}
\end{figure}

In \cref{fig:nlg_time_definitions} we illustrate these steps for two randomly selected BHs (one represented by a solid line in each panel and the other by a dashed line in each panel). The top panel shows the 50~Myr time-averaged sBHAR history, highlighting our starting point, the maximum value, $t_\mathrm{peak}$. The middle panel shows the gradient of the log$_{10}$\M{BH} history, highlighting our threshold value of d(log$_{10}$\M{BH})/d$t$ = 0.25 dex~Gyr$^{-1}$ as a horizontal dashed line. Where the histories first intersect with this threshold both backwards and forwards from the value $t_\mathrm{peak}$, defines $t_\mathrm{NLG[start]}$ and $t_\mathrm{NLG[end]}$ respectively. Finally, the bottom panel shows the BH mass history. Each line is colour coded via the identified phase of evolution: purple lines represent the stellar feedback regulated phase ($t < t_\mathrm{NLG[start]}$), orange lines the non-linear growth phase ($t_\mathrm{NLG[start]} \leq t \leq t_\mathrm{NLG[end]}$) and green lines the AGN feedback regulated phase ($t > t_\mathrm{NLG[end]}$).    

\subsection{Defining \ndyn: the most proximate merger}
\label{sect:def_ndyn}

To aid in establishing galaxy--galaxy mergers as potential triggering mechanisms for the non-linear phase in \cref{sect:mergers}, we introduce \ndyn, defined as the number of dynamical times between the start of the non-linear growth phase and the completion time of the most proximate (i.e., closest in time) merger, i.e.,

\begin{equation}
\label{eq:ndyn}
n_{\mathrm{dyn}} = \frac{t_{\mathrm{NLG[start]}} - t_{\mathrm{merger}}}{t_{\mathrm{dyn}}},
\end{equation}

\noindent where $t_{\mathrm{NLG[start]}}$ is the onset time of non-linear growth defined in \cref{sect:identifying_nlg}, $t_{\mathrm{merger}}$ is the completion time of the most proximate host galaxy merger and $t_{\mathrm{dyn}}$ is the dynamical time. We define the the dynamical time as the free-fall time of the dark matter halo, i.e.,

\begin{equation}
t_{\mathrm{dyn}} \equiv \left(\frac{3 \pi}{32 G (200
\rho_{\mathrm{crit}})}\right)^{1/2},
\end{equation}

\noindent where $\rho_{\mathrm{crit}}$ is the critical density of the Universe at $t_{\mathrm{NLG[start]}}$. For reference, $t_{\mathrm{dyn}} \approx$ 1.6~Gyr at $z=0$, $\approx$ 0.5~Gyr at $z=2$ and $\approx$ 0.2~Gyr at $z=5$. Thus negative (positive) values of \ndyn indicate that the most proximate merger completed after (before) the rapid growth phase began. We compute \ndyn separately for the most proximate major merger ($t_{\mathrm{merger}}$($\mu \geq \frac{1}{4}$), denoted \ndynx{maj}), minor merger ($t_{\mathrm{merger}}$($\frac{1}{10} \leq \mu < \frac{1}{4}$), denoted \ndynx{min}) and either a major or minor merger ($t_{\mathrm{merger}}$($\mu \geq \frac{1}{10}$), denoted \ndynx{all}). 

High values of \ndyn are capped to $\pm$10 dynamical times as mergers with $|n_{\mathrm{dyn}}| > 10$ are unlikely to have had an influence on the non-linear period. The BHs hosted in galaxies that did not experience any merger of a particular classification throughout their lifetime (and therefore have no valid value of \tmerger) are assigned the value \ndyn $=10$ to still contribute to the normalization of the merger rate.

\subsubsection{Creating a control sample of \ndyn}
\label{sect:control_ndyn}

To ascertain the significance of mergers in proximity to the non-linear phase, we require a control sample. Therefore for each BH's value of \ndynx{maj}, \ndynx{min} and \ndynx{all} we construct ten associated control values. These are obtained by recomputing \ndynx{maj}, \ndynx{min} and \ndynx{all} in ten random control galaxies using the $t_{\mathrm{NLG[start]}}$ value of the source galaxy (overriding the native value of $t_{\mathrm{NLG[start]}}$ in the control galaxies). The control galaxies are selected only on stellar mass (required to be within $\pm$0.5~dex of the source galaxy) and redshift, and therefore yield the expectation values of \ndynx{maj}, \ndynx{min} and \ndynx{all} that would be obtained for a galaxy of that mass, at that epoch, solely from the background merger rate, with no regard to the activity of the BH. For any collection of \ndyn values, such as the distributions in \cref{fig:nlg_ndyn_distribution}, we combine their associated control values to create ten control samples. Any deviations from the \ndyn distributions of the controls indicates the relative prevalence of mergers around the rapid growth phase over the background rate.

\section{Results}
\label{sect:results}
\subsection{Properties of the black holes}

\begin{figure}
\includegraphics[width=\columnwidth]{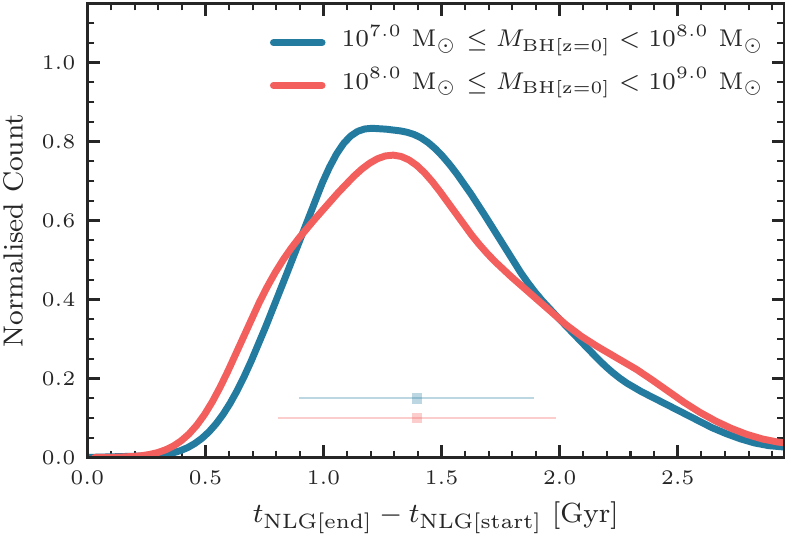}

\caption{The distribution of non-linear growth durations (i.e., $t_{\mathrm{NLG[end]}} - t_{\mathrm{NLG[start]}}$) for the BHs within our sample, separated into two present day BH mass ranges: $10^{7}$ \Msol $\leq M_{\mathrm{BH[z=0]}} < 10^{8}$ \Msol (red line) and $10^{8}$ \Msol $\leq M_{\mathrm{BH[z=0]}} < 10^{9}$ \Msol (blue line). The median values and 10$^{\mathrm{th}}$--90$^{\mathrm{th}}$ percentile ranges for each distribution are indicated by error bars ($1.4^{+0.6}_{-0.9}$~Gyr and $1.4^{+0.5}_{-0.7}$ for the upper and lower BH mass ranges, respectively). The median period of time BHs spend within the non-linear phase is insensitive to the eventual mass of the BH over this range.}

\label{fig:nlg_duration}
\end{figure}

We begin with investigating the properties of the BHs within our sample in relation to their rapid growth phase. \cref{fig:nlg_duration} shows the distribution of the non-linear phase durations (i.e., $t_{\mathrm{NLG[end]}} - t_{\mathrm{NLG[start]}}$), separated into two present day BH mass ranges: $10^{7}$ \Msol $\leq M_{\mathrm{BH[z=0]}} < 10^{8}$ \Msol (red line) and $10^{8}$ \Msol $\leq M_{\mathrm{BH[z=0]}} < 10^{9}$ \Msol (blue line). Both distributions are relatively narrow and broadly symmetric in their shape. The median duration of the rapid growth phase for the upper and lower present day BH mass ranges are almost identical ($1.4^{+0.6}_{-0.9}$~Gyr and $1.4^{+0.5}_{-0.7}$ respectively, the error values outline the 10$^{\mathrm{th}}$--90$^{\mathrm{th}}$ percentile ranges). Therefore the median period of time spent within the non-linear phase is insensitive to the eventual BH mass over this range. 

\begin{figure}
\includegraphics[width=\columnwidth]{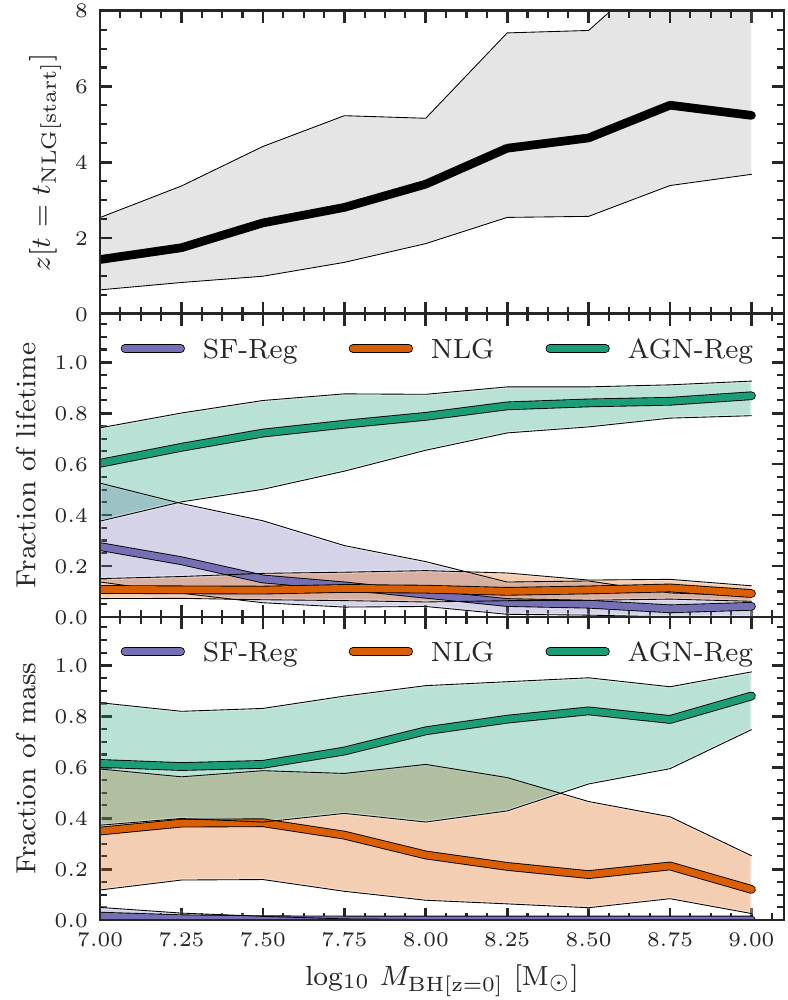}

\caption{Properties of the BHs within our sample in relation to their rapid growth phase. The solid lines are the median values and the shaded regions outline the 10$^{\mathrm{th}}$--90$^{\mathrm{th}}$ percentile ranges. Each property is plotted as a function of the final BH mass. \textit{Top panel}: the onset redshift of the rapid growth phase. \textit{Middle panel}: the fraction of the BHs lifetime that was spent in the three evolutionary phases. \textit{Bottom panel}: the fraction the total BH mass that was accumulated, via both mergers and accretion, in the three evolutionary phases.}

\label{fig:nlg_properties}
\end{figure}

Further properties of the rapid growth phase are investigated in \cref{fig:nlg_properties}. Here we show, from top to bottom, the onset redshift of the non-linear phase, the fraction of the BHs lifetime that was spent in the three evolutionary phases and the fraction of the total final BH mass that was accumulated, via both mergers and accretion, in the three evolutionary phases, each as a function of the final BH mass.

Starting with the top panel, we find today's most massive BHs began their non-linear phase, on average, the earliest ($z \approx 2$ for \M{BH[z=0]} $= 10^{7}$ \Msol increasing to $z \approx 6$ for \M{BH[z=0]} $= 10^{9}$ \Msol). This result is expected, as these BHs, which are hosted by some of the most massive haloes today \citep[e.g.,][]{McAlpine2017}, will tend to have reached the critical halo virial temperature for non-linear growth at earlier epochs than their lower mass counterparts. The fraction of a BHs lifetime spent in the rapid growth phase is low, and relatively constant for all the BHs within our sample ($\approx 15$\%, see middle panel). Most of the duration of massive BH life is spent in the AGN feedback regulated phase (between $\approx 60$ and $90$\% of their lifetimes). The fraction of the total BH mass that is accumulated in the non-linear phase is not constant; it accounts for $\approx 30$\% of the final mass for \M{BH[z=0]} $=10^{7}$ \Msol and decreases to $\approx 5$\% for \M{BH[z=0]} $=10^{9}$ \Msol (see bottom panel). Regardless of the time BHs spent in the stellar feedback regulated phase, which is only non-negligible for the lowest-mass BHs we study, almost no mass is accumulated, due to the quenching of BH growth via efficient stellar feedback. 

Therefore, the earlier BHs undergo their non-linear growth phase, the less contribution this phase has to the present day mass. Regardless of when this phase begins, it is generally short lived relative to the lifetime of the BH.

\subsubsection{Black hole activity during the rapid growth phase}

\begin{figure}
\includegraphics[width=\columnwidth]{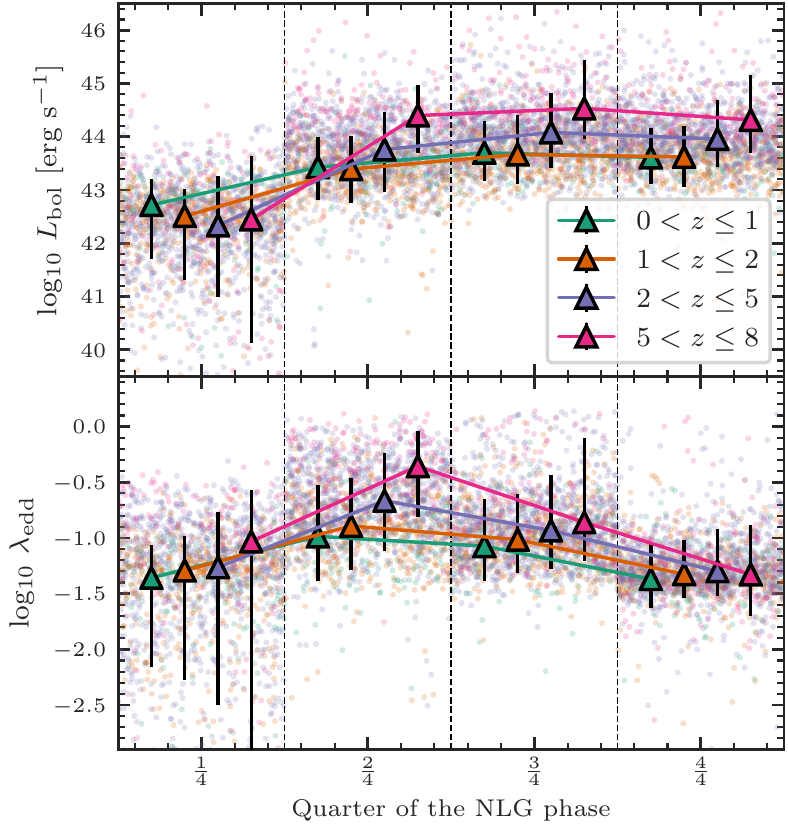}

\caption{The accretion activity of the BHs within our sample during their rapid growth phase. For each BH, the non-linear phase is divided into four equal time segments between $t_{\mathrm{NLG[start]}}$ and $t_{\mathrm{NLG[end]}}$, and the mean AGN luminosity (top panel) and the mean Eddington rate (bottom panel) is computed for each quarter. The solid circles are the mean values for each individual BH at a given quarter, coloured by the redshift at which they started their non-linear phase (i.e., $z[t = t_{\mathrm{NLG[start]}}]$), as indicated by the legend. We assign each BH a random scatter along the x-axis of each quarter bin, for clarity. The solid triangles indicate the median values of the four bins, with the error bars outlining the 10$^{\mathrm{th}}$--90$^{\mathrm{th}}$ percentile range. The median values are offset from each other along the x-axis, for clarity. The bolometric luminosity increases from the beginning to the end of the non-linear phase. The Eddington rate peaks at approximately 50\% of the way through the rapid growth phase. These trends are epoch independent, however at higher redshift the mean values increase in each property.}

\label{fig:quartiles}
\end{figure}

\begin{figure*}
\includegraphics[width=\textwidth]{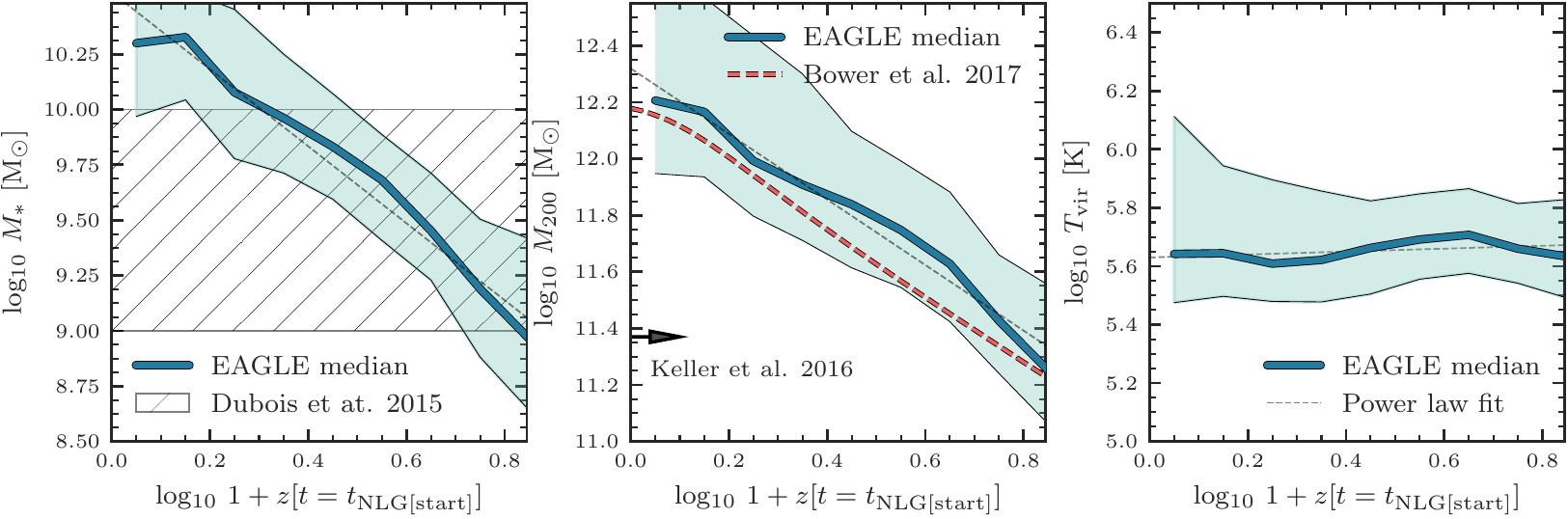}

\caption{The galaxy stellar mass (left panel), halo mass (middle panel) and halo virial temperature (right panel) of the hosts of the BHs within our sample at the beginning of their rapid growth phase ($t = t_{\mathrm{NLG[start]}}$) as a function of the redshift at which their rapid growth began. The solid lines indicate the median values, with the shaded regions outlining the 10$^{\mathrm{th}}$--90$^{\mathrm{th}}$ percentile ranges. Single power law fits to the median trends are indicated by dashed black lines. The BHs starting their rapid growth phase at low redshift do so in haloes and galaxies approximately an order of magnitude more massive than their high-redshift counterparts, indicating that there is no fixed halo or galaxy mass at which the rapid growth phase initiates, instead, BHs enter their rapid growth phase at a $\approx$ constant critical halo virial temperature ($T_{\mathrm{vir}} \approx 10^{5.6}$~K). Included in the left and middle panels are three theoretical predictions for the stellar/halo mass(es) at which stellar feedback can no longer efficiently regulate the gas content of the galaxy, and stalls, marking the theoretical transition point to the non-linear phase of BH growth, see \cref{sect:disc_failingfeedback} for a discussion.}

\label{fig:mass_at_nlg}
\end{figure*}

The accretion activity of the BHs within our sample during their rapid growth phase is investigated in \cref{fig:quartiles}. For each BH, we divide the non-linear phase into four equal time segments\footnote{Note that the absolute time intervals of the quarters will be different for each BH due to the varying range of non-linear growth durations (see \cref{fig:nlg_duration}).} between $t_{\mathrm{NLG[start]}}$ and $t_{\mathrm{NLG[end]}}$ and measure the mean bolometric AGN luminosity ($L_{\mathrm{AGN}}$\footnote{Defined as $L_{\mathrm{AGN}} = \epsilon_{r} \dot M_{\mathrm{BH}} c^{2}$, where $\epsilon_{r}$ is the radiative efficiency of the accretion disk, which is assumed to be 0.1
\citep{Shakura1973}.}, top panel) and the mean Eddington rate ($\lambda_{\mathrm{edd}}$\footnote{Defined as $\lambda_{\mathrm{edd}} = \dot M_{\mathrm{BH}} / \dot M_{\mathrm{edd}}$ where $\dot M_{\mathrm{BH}}$ is the accretion rate of the BH and $\dot M_{\mathrm{edd}}$ is the Eddington limit. The BH accretion rate in the \eagle reference model is capped to the Eddington limit over $h$ (i.e., the maximum allowed value of $\lambda_{\mathrm{edd}} = 1/h = 1.48$).}, bottom panel) for each quarter. This allows us to measure the comparative trends of BH activity throughout each segment of the rapid growth phase. The BHs are separated by the redshift at which they began their non-linear phase (i.e., $z[t = t_{\mathrm{NLG[start]}}]$). 

The general evolutionary trend for both the AGN luminosity and the Eddington rate through the non-linear phase is very similar for each redshift range. The AGN luminosity in the $1^{\mathrm{st}}$ quarter initiates at a relatively low rate ($\sim 10^{42}$ \erg), steadily increases towards the $3^{\mathrm{rd}}$ quarter ($\sim 10^{44}$ \erg) and remains approximately at this level through to the $4^{\mathrm{th}}$ quarter. This behavior is consistent with the scenario of a growing BH embedded within a relatively constant source of fuel. The Eddington rate similarly begins at a relatively low level in the $1^{\mathrm{st}}$ quarter ($\lambda_{\mathrm{edd}} \sim 10^{-1.5}$), evolves towards a peak in the $2^{\mathrm{nd}}$ and $3^{\mathrm{rd}}$ quarters ($\lambda_{\mathrm{edd}} \sim 10^{-0.5}$), and finally reduces back to values similar to that of the $1^{\mathrm{st}}$ quarter. This remains consistent with the picture seen in the panel above: the AGN luminosity of a growing BH tends to a constant rate in the later states of non-linear growth. For each of the two properties, the mean values increase with increasing redshift, indicating that the BHs that underwent their rapid growth phase at higher redshift are on average more luminous and closer to the Eddington limit than their counterparts at lower redshift. If we examine the individual mean Eddington rate values (background coloured circles), we find that it is extremely rare to sustain continued growth at the Eddington limit for any period during the non-linear phase.

\subsection{Properties of the hosts at the start of the rapid growth phase}

We now turn to the galaxies and dark matter haloes that host the BHs within our sample at the onset of their rapid growth phase. \cref{fig:mass_at_nlg} shows, from left to right, the galaxy stellar mass, halo mass and halo virial temperature\footnote{The virial temperature of the halo is defined as $T_{\mathrm{vir}} = \mu m_{\mathrm{p}} V_{\mathrm{c}}^{2}/ 5 k_{\mathrm{b}}$, where $\mu$ is the mean molecular weight of the gas in the halo (assumed to be 0.59 for a primordial gas), $m_{\mathrm{p}}$ is the mass of the proton, $k_{\mathrm{b}}$ is the Boltzman constant and $V_{\mathrm{c}} = G M_{200} / r_{200}$ is the virial velocity \citep{MoBoschWhite}.}, each at the time $t = t_{\mathrm{NLG[start]}}$, as a function of the redshift at which the rapid growth phase began. There is a distinct negative trend visible in the first two panels, with both the host galaxy and halo mass decreasing as the redshift increases (\M{*} $\approx 10^{10.5}$ \Msol at $z \approx 0$ decreasing to \M{*} $\approx 10^{9}$ \Msol at $z \approx 6$ and \M{200} $\approx 10^{12.4}$ \Msol at $z \approx 0$ decreasing to \M{200} $\approx 10^{11.2}$ \Msol at $z \approx 6$). There appears, therefore, to be no fixed galaxy or halo mass at which non-linear BH growth initiates, instead, the rapid growth phase of BHs appears to ubiquitously initiate when the host halo reaches a critical virial temperature ($T_{\mathrm{vir}} \approx 10^{5.6}$~K, see right panel). This is consistent with the physical scenario outlined in \cref{sect:how_bhs_grow}, whereby the buoyancy of the stellar feedback driven outflows stall at a critical halo virial temperature, allowing the gas density within the center of the galaxy to rise, triggering the rapid growth phase.    


We fit each of the median trends with a single power law relation using the python module {\sc lmfit}\footnote{https://lmfit.github.io/lmfit-py/}, indicated on the figure as dashed black lines. The $1 \sigma$ errors on the median values inserted into the fitting routine are computed from bootstrap resampling. The fits are: log$_{10}(M_{*}/M_\odot) = (-1.74 \pm 0.11)\mathrm{log}_{10}(1+z) + 10.53 \pm 0.06$, log$_{10}(M_{200} / M_{\odot}) = (-1.16 \pm 0.07)\mathrm{log}_{10}(1+z) + 12.32 \pm 0.04$ and log$_{10}(T_{\mathrm{vir}} / K) = (0.05 \pm 0.04)\mathrm{log}_{10}(1+z) + 5.63 \pm 0.02$, from the left to right panels, respectively.

\subsection{The proximity of mergers to the rapid growth phase}
\label{sect:mergers}

\begin{figure} \includegraphics[width=\columnwidth]{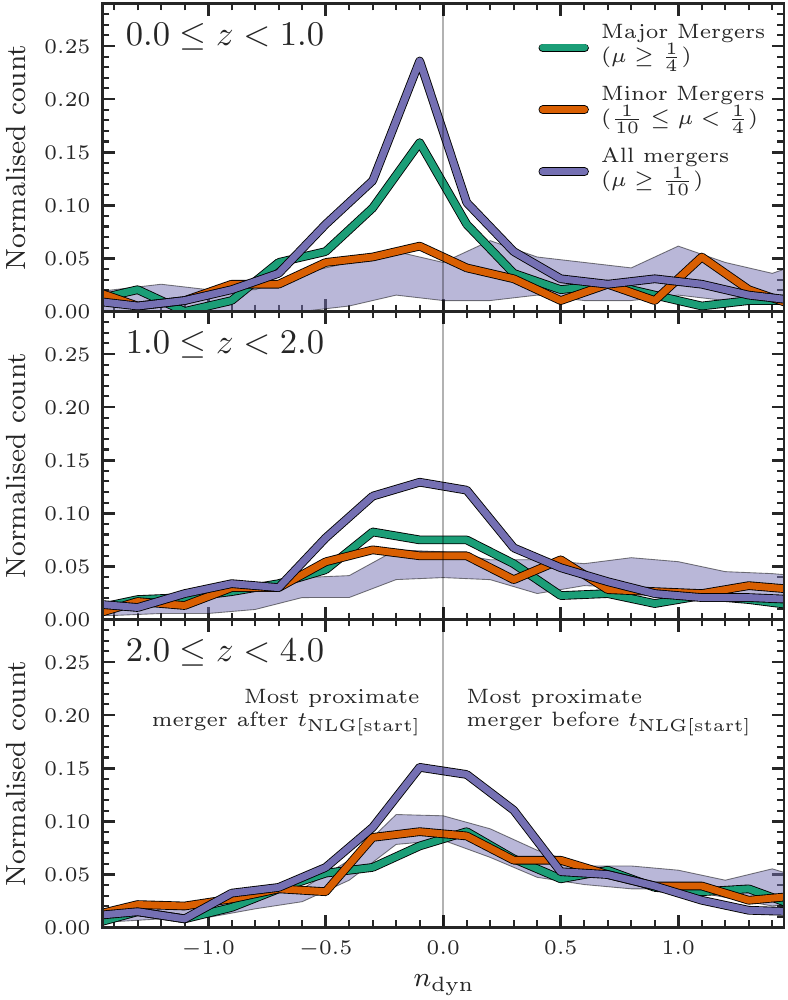}
\caption{The distributions of \ndynx{maj} (green lines), \ndynx{min} (orange lines) and \ndynx{all} (purple lines) for each BH contained within our sample (the number of dynamical times since the most proximate in time merger, see \cref{sect:def_ndyn} for definitions). The BHs are separated into those that began their rapid growth phase in the redshift ranges $0.0 \leq z < 1.0$ (top panel), $1.0 \leq z < 2.0$ (middle panel) and $2.0 \leq z < 4.0$ (bottom panel). The shaded regions outline the 10$^{\mathrm{th}}$--90$^{\mathrm{th}}$ percentile range of the control distributions for \ndynx{all} (see \cref{sect:control_ndyn}). These reveal the predicted distribution of \ndynx{all} that would be produced solely from the background merger rate. The distributions are normalized by the total number of BHs in that redshift range, including those with host galaxies that experienced no mergers of the particular classification in their lifetimes (see \cref{sect:def_ndyn}). The significant peak just before the value \ndynx{all} $=0$ in the upper panel, relative to the control, shows that mergers commonly trigger this phase of BH evolution at low redshift (almost exclusively from major mergers). At higher redshifts the peak lowers and the distribution broadens, with the distributions falling closer to that of the control sample.}

\label{fig:nlg_ndyn_distribution} \end{figure}

\begin{figure} \includegraphics[width=\columnwidth]{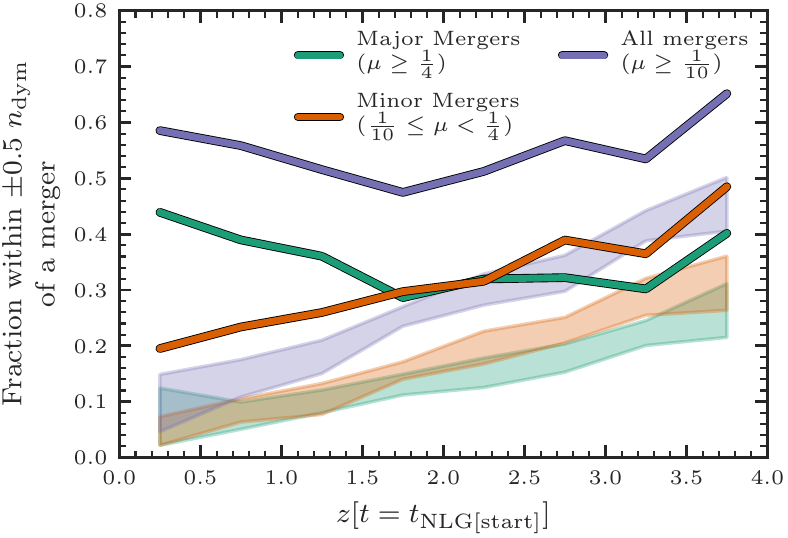}
\caption{The fraction of BHs within our sample that began their rapid growth phase within $\pm$0.5 dynamical times of a major merger (green line), minor merger (orange line) and either a minor or major merger (purple line) as a function of the redshift at which the rapid growth phase began. The fractions that would be expected from the background merger rate for similar mass galaxies (with no regard for BH activity) are shown as shaded regions. A substantial excess of BHs at low redshift are more proximate in time to a merger than the control prediction. Therefore mergers, almost exclusively major mergers, are triggering the rapid growth phase for a large fraction of the BHs at lower redshift.}

\label{fig:nlg_ndyn_fraction} \end{figure}

We conclude this section by investigating the physical connection between the start of the non-linear phase of BH growth and galaxy mergers. \cref{fig:nlg_ndyn_distribution} shows the distributions of \ndynx{maj} (green lines), \ndynx{min} (orange lines) and \ndynx{all} (purple lines) for each BH contained within our sample (see \cref{sect:def_ndyn} for their definitions). The BHs are separated into those that began their rapid growth phase in the redshift ranges $0.0 \leq z < 1.0$, $1.0 \leq z < 2.0$ and $2.0 \leq z < 4.0$\footnote{We note that whilst there are galaxies that begin their non-linear phase at $z > 4$ (see \cref{fig:nlg_properties}), we limit our merger analysis to $z < 4$ to ensure we adequately resolve minor mergers ($M_{*,2} \geq 10^{8}$ \Msol, see \cref{sect:method}) for all galaxies, as the host galaxies of the BHs beginning their rapid growth at $z < 4$ have masses $M_{*} \geq 10^{9}$ \Msol (see \cref{fig:mass_at_nlg}).}, shown in the top, middle and bottom panels, respectively. These distributions reveal the characteristic proximity in time between galaxy--galaxy mergers of the host and the onset of the rapid growth phase of the central BH.

Starting with the top panel, we find that the distribution of \ndynx{all} (purple line) strongly peaks just before the value \ndynx{all} $=0$ (indicated by a vertical black line). The abundance of quantitatively low values of $|n_{\mathrm{dyn[all]}}|$ indicates  that for these BHs there is often either a major or minor merger during this phase of their evolution. Additionally, the preference for negative values tells us that the most proximate merger generally completes \textit{after} the non-linear phase has begun. If we were to attribute the most proximate merger as the triggering mechanism, it would indicate that the rapid growth phase initiates during the initial period of the interaction and well before the final coalescence of the two galaxies. If we consider minor and major mergers independently (orange and green lines), we find that most of the peak counts for all mergers is contributed by major mergers, rather than minor mergers. As we move to higher redshifts, in the middle and bottom panels, we find the distribution broadens and the peak lowers and shifts closer to the value \ndyn $\approx 0$.     

However, it is difficult to attribute any significance to these peaks without also knowing the expected distribution of \ndynx{min}, \ndynx{maj} and \ndynx{all} that would arise just from the background merger rate, regardless of BH activity. For this we additionally show the 10$^{\mathrm{th}}$--90$^{\mathrm{th}}$ percentile range of the ten control samples (see \cref{sect:control_ndyn}) for \ndynx{all} as a shaded region in each panel. For clarity, we exclude the control samples for the remaining two merger classifications from this figure, but note that they are indistinguishable from the control distribution that is plotted. Relative to the control, it is clear that the enhancement around the value \ndynx{all} $\approx 0$ is a significant deviation from what is expected from the background rate, particularly at low redshift.

To measure this enhancement more clearly, we present \cref{fig:nlg_ndyn_fraction}. This shows the fraction of BHs that began their non-linear phase within $\pm$0.5 dynamical times of a merger as a function of the redshift at which the non-linear phase began for the same three merger classifications. We additionally show the predicted fractions from our control sets as shaded regions. The behavior first hinted towards in \cref{fig:nlg_ndyn_distribution} is now much more apparent. There is always an excess above the control, indicating that mergers are more common around the start of the rapid growth phase than one would predict from the background rate. At low redshift ($z \approx 0$) the excess is substantial; $\approx 60$\% of the BHs starting their rapid growth phase at this time are found to be within $\pm$0.5 dynamical times of either a minor or major merger, when only $\approx 10$\% would be expected to be so from the background rate. It therefore appears that mergers, primarily major mergers, are strong drivers of the rapid growth phase for many BHs at low redshift. We discuss this result further in \cref{sect:disc_mergers}. 

\section{Discussion}
\label{sect:discussion}
\subsection{Stalling stellar feedback and the transition to the rapid growth phase of black holes}
\label{sect:disc_failingfeedback}

Whilst a number of current hydrodynamical simulations have reported a link between efficient stellar feedback and the substantial reduction of BH growth in low-mass systems, it remains unclear exactly when, and how, the transition between stalling stellar feedback and the onset of rapid BH growth occurs. 

\citet{Dubois2015} study the growth of an individual dark matter halo ($10^{12}$ \Msol at $z=2$) by means of a high-resolution cosmological zoom in, taken from the {\sc seth} simulation suite using the adaptive mesh refinement code {\sc ramses} \citep{Teyssier2002}. They find that at redshifts $z > 3.5$ the galaxy's central reservoir of gas is sufficiently disrupted via efficient stellar feedback\footnote{This is only true when their delayed cooling prescription for stellar feedback is used \citep{Teyssier2013}. When performing similar tests with a kinetic stellar feedback model \citep{Dubois2008}, they only find a very limited effect on the growth of the central BH.} to substantially restrict the accretion onto the central BH. After the galaxy has accumulated sufficient mass, they witness a decline in the ability of stellar feedback to disrupt the gas, and the central BH transitions into a rapid growth phase. They argue that this transition is directly linked to the balance between the momentum-driven stellar wind and the escape velocity of the central bulge. From this they predict the theoretical mass scale above which these winds can no longer escape the bulge, leading to a rise in the central gas density, which in turn feeds the central BH. They state the escape velocity for a bulge of mass $10^{9}$ \Msol at a fixed bulge radius of 100~pc is $\approx 270$~km s$^{-1}$, approximately equal to that achieved by a supernovae Sedov blast wave (see their Equation 1). This is indeed the bulge mass found by their simulation around which the rapid BH growth begins. The prediction that stellar feedback begins to stall ubiquitously at a constant bulge mass and bulge radius (i.e., a constant density) implies the existence of a critical mass that is independent of epoch, contrary to our findings in \cref{fig:mass_at_nlg}. We show this bulge mass (converted to a range of total stellar masses assuming a bulge to total stellar mass ratio of between 0.1 and 1.0) as a hatched region in the left panel of \cref{fig:mass_at_nlg}. We note that, from the study of a single halo, one cannot capture the variation of the critical mass with time and halo properties. Indeed, the assumption of a fixed bulge density is potentially a key assumption that leads to a redshift-independent critical mass, though we know that bulges at high redshift are denser than those in the local Universe.

\citet{Keller2016} investigated the evolution of 18 isolated Milky Way-like disc galaxies from the MUGS2 simulation suite \citep{Stinson2010}, performed using the smoothed particle hydrodynamics code {\sc gasoline2} \citep{Wadsley2017}. They find that supernovae alone cannot regulate the incoming gas flow to systems with virial masses $> 10^{12}$ \Msol, which can result in a runaway production of stars in the central bulge. This stalling is attributed to the shutdown of galactic winds from a deepening potential well (mass loading factors fall from a relatively constant level of $\eta \sim 10$ below the critical mass scale, to $\eta < 1$ just above). They report a redshift-independent central baryonic mass of $10^{10.0 \pm 0.1}$ \Msol and  halo mass of $10^{11.37 \pm 0.08}$ \Msol at which the stellar feedback begins to stall. This halo mass is indicated by an arrow in the middle panel of \cref{fig:mass_at_nlg}. Although a universal and non-evolving critical mass is again in conflict with our findings (see \cref{fig:mass_at_nlg}), we note that only a moderate range of present day galaxy masses are explored in the simulation set of \citet{Keller2016} ($M_{\mathrm{*[z=0]}} = $0.5-20.8$\times 10^{10}$ \Msol). Furthermore, there is evidence of a varying critical halo mass even within this limited mass range (see their Figures 7 \& 8). Perhaps most importantly, as no prescription for BHs is included for these simulations, they are unable to directly investigate the link between stalling stellar feedback and the rapid growth phase. The runaway production of stars seen in systems above this critical mass, however, strongly suggests that AGN feedback (and thus a massive BH) is required to curb continued galaxy growth.   

\citet{AnglesAlcazar2017} perform a set of high-resolution cosmological hydrodynamic simulations of quasar-mass halos ($M_{\mathrm{halo}}(z = 2) \approx 10^{12.5}$~\Msol) using the {\sc fire-2} simulation code \citep{Hopkins2018}. These simulations model stellar feedback by supernovae, stellar winds, and radiation, and BH growth using a gravitational torque-based
prescription (see also \cref{sect:dependence_on_model}), however no AGN feedback is implemented. They discover that early BH growth in low mass galaxies is extremely limited by bursty stellar feedback continuously evacuating gas from the galactic nuclei ($\lesssim 100$~pc). The BHs during this time remain significantly undermassive relative to their galaxy host. Only as the galaxy approaches $M_{\mathrm{bulge}} \sim 10^{10}$~\Msol does BH growth start to become more efficient, as the nuclear stellar potential begins to retain a significant gas reservoir, and the star formation becomes less bursty. In this more massive regime, the BHs are then seen to rapidly converge onto the \M{BH}--\M{bulge} scaling relation. Analogous to \citet{Dubois2015}, they attribute this transition to the increased escape velocity of the bulge now exceeding that of the stellar feedback-driven winds, and also suggest the possibility of a redshift independent critical mass. 

\citet{Bower2017} provide a different explanation. They develop a simple analytical model that describes the interaction between buoyant, high entropy star formation driven outflows and the rate of the cosmic gas inflow. In low mass systems ($M_{\mathrm{200}} \lesssim 10^{12}$ \Msol) the adiabat of this outflow exceeds that of the haloes diffuse corona, and can buoyantly escape. This ensures that the central gas densities within the galaxy remains low, and the central BH is deprived fuel. In massive systems a hot corona forms, and the star formation-driven outflows are no longer buoyant relative to their surroundings. This triggers a high density build up of gas within the central regions of the galaxy, and a subsequent non-linear response from the central BH. The critical halo mass predicted for this transition is given by their Equation 5, which we show in the middle panel of \cref{fig:mass_at_nlg} as a red dashed line. There is a good agreement between the analytical prediction and that of our findings, reproducing the redshift dependence, with only a small offset in the normalization between the two trends. We note, that whilst the model of \citet{Bower2017} was validated against the \eagle simulation, it was independently derived, and not calibrated using the simulation results.

To summarize, we find that the critical galaxy/halo mass at which stellar feedback stalls and rapid BH growth begins is not constant, and decreases with increasing redshift. Instead, we find that rapid BH growth phase initiates at an approximately constant halo virial temperature (see \cref{fig:mass_at_nlg}). This is contrary to some previous predictions, where an epoch-independent single critical mass has been reported. But, we understand this as limitations of these works due to a limited range of simulated parameters, or because AGN feedback was not included in these simulations.  

\subsection{The role of galaxy mergers in triggering the rapid growth phase of black holes}
\label{sect:disc_mergers}

In the paradigms set out by the studies in the previous section, the primary factor in transitioning from efficient to inefficient stellar feedback-driven outflows is the secular evolution of the bulge/galaxy/halo. That is, when the host system becomes sufficiently massive, the stellar winds/outflows become trapped via a deepening potential well or hot corona. However, the rapid growth phase of BHs may also, or exclusively, be triggered by galaxy--galaxy interactions.

\citet{Dubois2015} found for the evolution of a single halo (discussed in the previous section) that the rapid growth phase of the central BH was likely triggered by a major merger. In \cref{sect:mergers} we found a strong connection between the onset time of non-linear growth ($t = t_\mathrm{NLG[start]}$) and the most proximate merger, regardless of the redshift at which non-linear growth began. Approximately 60\% of the BHs within our sample initiated their rapid growth phase within $\pm$$0.5$ dynamical times of either a minor or major merger ($> 40$\% a major merger, see \cref{fig:nlg_ndyn_fraction}). At lower redshifts ($z \approx 0$), the merger fractions far exceeded the predicted proximity to mergers from the background rate ($\approx 60$\% versus $\approx 10$\%), whereas at higher redshifts ($z \approx 4$), the merger fractions fell much closer to the predicted rate \citep[$\approx 60$\% versus $\approx 45$\%, as the background merger rate increases with increasing redshift][]{Rodriguez-Gomez2015,Qu2017}. We could interpret this in two ways: (1) a galaxy's central BH at lower redshift increasingly \textit{requires} a major disturbance to initiate its rapid growth phase, derived from the increasing excess in the merger fractions above the control sample, or (2) mergers are always important for triggering the rapid growth phase, derived from the universally high merger fractions, and the fact that all galaxies ubiquitously experience mergers more frequently at higher redshifts is inconsequential.

In either case, galaxy interactions appear to be important triggering mechanisms for the rapid growth phase, at least in the low redshift Universe. From this one may conclude that mergers can act as catalysts to accelerate the transition from stalling stellar feedback to the rapid growth phase, however, the relatively low spread in halo virial temperatures at which the rapid growth phase initiates would suggest that this is not the case (see right panel of \cref{fig:mass_at_nlg}). It appears, then, that whilst the non-linear phase may be initiated through a strong interaction, a characteristic halo virial temperature remains essential for rapid BH growth to occur. 

\subsection{Observing the rapid growth phase of black holes}

\begin{figure}
\includegraphics[width=\columnwidth]{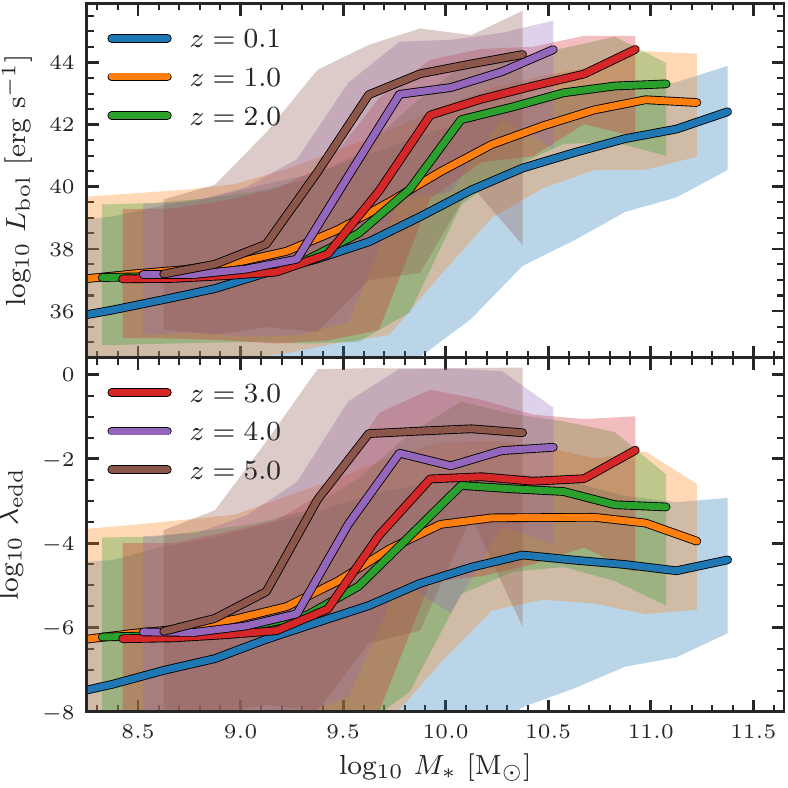}

\caption{The median bolometric AGN luminosity (upper panel) and the median Eddington rate (lower panel) as a function of the host galaxy stellar mass for six redshifts, as indicated in the legend. These are computed from all galaxies at the stated epoch, and not only those hosting the BHs contained within the massive BH sample outlined in \cref{sect:sample_selection}. The shaded regions outline the  10$^{\mathrm{th}}$--90$^{\mathrm{th}}$ percentile range.}

\label{fig:L_vs_SM}
\end{figure}

We explore the considerations needed to validate the non-linear phase in observations of the BH population in \cref{fig:L_vs_SM}. This figure shows the median bolometric AGN luminosity (top panel) and the median Eddington rate (bottom panel) for all the BHs within the \eagle volume as a function of the host galaxy stellar mass at six different redshifts. Here we see the familiar imprint of the three phases of BH evolution: before the critical halo virial temperature BHs are effectively inactive, the luminosities and Eddington rates then increase by many orders of magnitude over a narrow stellar mass window around the critical halo virial temperature, and finally the luminosities and Eddington rates come to settle to an approximately constant median rate after the critical halo virial temperature, though with very large scatters. As we saw in \cref{fig:mass_at_nlg}, the critical \emph{mass} marking this transition reduces with increasing redshift. \cref{fig:L_vs_SM} also shows that the increase in AGN luminosity and Eddington rate during the non-linear phase is larger at high redshift.

One could then in principle observe evidence of the rapid growth phase in two ways: attempt to discover the transition between inactive BHs and moderately active BHs in low-mass galaxies, or find the transition between a steep and shallow relationship for the median $L_{\mathrm{bol}}$ and $\lambda_{\mathrm{edd}}$ around the critical halo virial temperature. The pivot \emph{mass} in each case is predicted to decrease as the redshift increases. However, the spread of many orders of magnitude in the AGN luminosity (the shaded regions outline the 10$^{\mathrm{th}}$--90$^{\mathrm{th}}$ percentile range), the difficulty in detecting low luminosity AGN ($L_{\mathrm{bol}} < 10^{43}$ \erg), the relatively narrow range and therefore the need for accurate measurements of the stellar masses, and the need for large statistical samples of objects at multiple epochs will make this extremely challenging. It is therefore more plausible to find evidence for the rapid growth phase \textit{indirectly} via the integrated BH accretion rate, i.e., the BH mass, as the three phases of BH evolution are also present within the BH mass--stellar mass relation \citep[e.g.,][ for the case of \eagle]{Crain2015,Schaye2015,Barber2016,RosasGuevara2016,Bower2017,McAlpine2017}. The scatter in this relation is also predicted to change considerably with the mass of the galaxy host: galaxies below the critical halo virial temperature will host BHs with a small scatter around the seed mass; galaxies around the critical halo virial temperature will host a large dynamic range of BH masses, due to the rapid BH growth over this mass range; and BHs hosted in galaxies above the critical halo virial temperature return to a much smaller scatter due to the regulation from AGN feedback. Indeed, changing  relationships between the mass of the galaxy host and that of the central BH across a range of stellar masses and morphologies have been found by empirical studies \citep[e.g.,][]{Scott2013,Greene2016,Lasker2016,Martin-Navarro2018}.   

\subsection{The dependence on the model}
\label{sect:dependence_on_model}

Three astrophysical prescriptions are crucial for forming the three phases of BH evolution investigated by this study: efficient stellar and AGN feedback, capable of regulating the gas inflow onto low- and high-mass galaxies, respectively, and the ability for BHs to grow rapidly when neither of these feedback processes are dominant. It is interesting to ask, then, to what extent the models that govern these processes influence the behavior of BH growth in hydrodynamical simulations, and how ubiquitous the creation of these three phases may be.

Efficient stellar feedback using many different model implementations across a range of resolutions is found to restrict the growth of BHs within low-mass galaxies in hydrodynamical simulations (see \cref{sect:disc_failingfeedback}). Interestingly, the {\sc illustris} project \citep{Vogelsberger2014}, which is a cosmological hydrodynamical simulation that shares many similarities with the \eagle project, shows no strong evidence of such behavior \citep{Sijacki2015}. Unlike \eagle, which models stellar feedback purely via the thermal injection of energy, {\sc illustris} adopt a kinetic wind model that temporarily decouples the hydrodynamics. Kinetic injection schemes can be less efficient at disrupting early BH growth \citep[see][ for example]{Dubois2015}. However, in the updated {\sc illustris-tng} model \citep{Pillepich2018}, where stellar feedback is now implemented partially thermally with a deliberate increased efficiency towards higher redshifts and in low-mass haloes, BH growth now appears limited in the familiar fashion below a critical mass \citep{Pillepich2018,Weinberger2018}. This phase of BH evolution is undoubtedly sensitive to the efficiency of the chosen stellar feedback model, however, efficient stellar feedback is crucial for replicating many of the observed properties of galaxies in hydrodynamical simulations, such as their sizes and star formation rates, and many hydrodynamical simulations have converged towards implementing a form of efficient stellar feedback as a result. Observations of BH activity (or lack thereof) in low-mass galaxies may therefore provide key insight for constraining stellar feedback models.  

The choice of BH growth model may also have interesting implications. Many of the widely used and successful BH growth models that have faithfully replicated many of the observed properties of BHs in the local Universe are derived from the original Bondi prescription \citep{Bondi1944}, which is also the basis for the BH growth model within the \eagle simulation \citep{RosasGuevara2015}. As Bondi-like accretion is proportional to the mass of the BH squared, BHs have the opportunity to grow at a rapid, non-linear rate if the conditions are favorable, hence the origin of such a short-lived rapid growth phase found by this study (see \cref{fig:nlg_duration}). However, there are other BH growth models with alternate dependences on the mass of the BH, such as the gravitational torque-based prescription introduced by \citet{Hopkins2011}, for which the accretion rate is proportional to the mass of the BH to a much lower power ($\frac{1}{6}$). In this regime, BHs do retain the capability to \squotes{rapidly} grow, as is shown by \citet{AnglesAlcazar2017}, however at a sub-Bondi rate. This would presumably lengthen the duration of the rapid growth phase, yet once the BH becomes sufficiently massive it would still enter the AGN feedback regulated phase, and the three phases of BH evolution would theoretically remain distinct. Additionally, BH growth models can be sensitive to the resolution and scale over which the accretion rate is estimated \citep[see][]{AnglesAlcazar2017}, which may also impact the result.  Observational measurements of the (changing) behavior of the \M{BH}--\M{*} relation around and beyond the critical transition mass will provide useful constraints between the different BH growth models.

Ultimately, to fully disentangle the direct influence of the stellar feedback and BH growth models on the three phases of BH evolution will require a parameter exploration coupled to a similar investigation as performed in this study.

\section{Conclusions}
\label{sect:conclusions}

We have investigated the rapid growth phase of BHs using the hydrodynamical cosmological \eagle simulation. Our main conclusions are as follows:

\begin{itemize}

\item \textbf{The majority of massive BH life is spent in the AGN feedback regulated phase, at $\approx$60--90\% of their lifetime.} The median duration of the rapid growth phase is only $\approx 1.4$~Gyr, corresponding to $\approx 15$\% of their lifetime. The fraction of the present day BH mass accumulated during the rapid growth phase decreases with increasing BH mass ($\approx 30$\% at \M{BH[z=0]} $= 10^{7}$ \Msol, decreasing to $\approx 5$\% at \M{BH[z=0]} $= 10^{9}$ \Msol). The remainder is acquired during the AGN feedback regulated phase, as no significant BH growth occurs during the stellar feedback regulated phase. See \cref{fig:nlg_duration,fig:nlg_properties}.  

\item \textbf{BHs enter the rapid growth phase at a critical halo virial temperature ($T_{\mathrm{vir}} \approx 10^{5.6}$~K).} There is no fixed host galaxy stellar mass or halo mass at which the rapid growth phase begins. BHs initiating their rapid growth phase today do so in galaxies and haloes approximately an order of magnitude more massive than their high-redshift counterparts (\M{*} $\approx 10^{10.5}$ \Msol at $z \approx 0$ decreasing to \M{*} $\approx 10^{9}$ \Msol at $z \approx 6$ and \M{200} $\approx 10^{12.4}$ \Msol at $z \approx 0$ decreasing to \M{200} $\approx 10^{11.2}$ \Msol at $z \approx 6$). See \cref{fig:mass_at_nlg}.

\item \textbf{Galaxy--galaxy interactions are important for triggering the rapid growth phase.} Approximately 60\% of the BHs initiating their rapid growth phase today ($z \approx 0$) do so within $\pm 0.5$ dynamical times of either a minor or major galaxy--galaxy merger ($\mu \geq \frac{1}{10}$) and $\approx 40$\% do so within $\pm 0.5$ dynamical times of a major merger ($\mu \geq \frac{1}{4}$). This is substantially higher than what is predicted from the background merger rate ($\approx 10$\%). At higher redshifts the merger fractions remain high ($\approx$ 60\%), however the background merger rate has also substantially increased by these epochs ($\approx$ 45\%), making it difficult to directly disentangle the importance of mergers in triggering the rapid growth phase at high redshift. Minor mergers play much less of a role in triggering the rapid growth phase at all epochs. See \cref{fig:nlg_ndyn_distribution,fig:nlg_ndyn_fraction}.



\end{itemize}

\section*{Acknowledgements}

This work was supported by the Science and Technology Facilities Council (grant number ST/P000541/1).

This work used the DiRAC Data Centric system at Durham University, 
operated by the Institute for Computational Cosmology on behalf of the 
STFC DiRAC HPC Facility (\url{www.dirac.ac.uk}). This equipment was funded 
by BIS National E-infrastructure capital grant ST/K00042X/1, STFC capital 
grant ST/H008519/1, and STFC DiRAC Operations grant ST/K003267/1 and 
Durham University. DiRAC is part of the National E-Infrastructure.

RAC is a Royal Society University Research Fellow.